\documentstyle [12pt]{article}

\begin{document}

\vspace{10mm}
\centerline{\Large\bf A simple formula}
\centerline{\Large\bf for Bose-Einstein corrections}
\vspace*{1cm}
\begin{center}
{\large J. Wosiek} 
\\
\vspace*{1.5mm}
Institute of Physics, Jagellonian University \\
Reymonta 4, 30-059 Cracow, Poland 
\end{center}
\begin{abstract}
In analogy with the quantum field theory of free bosons
a simple integral representation is derived for recently
proposed corrections describing the Bose Einstein effect.
The saddle point approximation to these integrals results
in a compact expression which sums effectively the original
$n!$ terms with an accuracy better than 2\% for $n > 7$
strongly correlated bosons.  
\end{abstract}
\vspace*{0.5cm}
 \vspace*{3.5cm} \newline
TPJU-1/97 \newline
January 1997 \newline
hep-ph/9701379 \newline 
\newpage
 
\section{Introduction}
   The very nature of the Bose-Einstein correlations makes them
rather difficult to include in the Monte Carlo event generators and
consequently 
the problem has been  the area of intensive studies for many years
 \cite{HBT} -\cite{ZJ}.
Recently Bia\l as and Krzywicki have proposed a simple recipe
to effectively include Bose-Einstein correlations in the 
event generating Monte Carlo programs \cite{BK}. 
The basic step in their procedure
consists of calculating a positive weight for each generated
event
\begin{equation}
W_n(p_1,...,p_n)=\sum_{{\cal Q}_n} \prod_i^n A_{i{\cal Q}_n(i)}=perm(A), 
\label{wei}
\end{equation}
where $A$ is a correlation matrix 
which depends on particle momenta
and on the choice of variables parameterizing the whole phenomenon.
One commonly used form reads
\begin{equation}
A_{ik}=\exp{\left(-{(p_i-p_k)^2\over 2\sigma^2}\right)}.  \label{cor}
\end{equation}
The sum in Eq.(\ref{wei}) extends over all permutations, ${\cal Q}_n$,
of $n$ elements (1,2,\dots ,n). Hence the numerical cost to calculate 
above weights
grows like $n!$ and this limits practical applications.

In this letter we derive a simple $2n$-dimensional integral representation
for a general permanent of a correlation matrix, $perm(A)$. Our representation
has two advantages. First, the integrals can be done analytically, in the
saddle point approximation, providing relatively simple and accurate
expression for the sum (\ref{wei}). Secondly, the integral representation
has a straightforward probabilistic interpretation. This indicates
that implementation of the correction scheme proposed in Ref.\cite{BK}
might be possible stochastically, and may advance the novel generation 
of MC programs.
 
     The motivation for this work came from the simple observation
that even the free quantum field theory of bosons automatically incorporates
symmetrization. It is therefore
plausible that the simple integral representation, analogous to that for
the propagation of $n$ identical bosons, should exist for the sum 
(\ref{wei}). In the next Setion we prove that this is indeed the case.  
\section{Integral representation}
   Let us recall the Wick theorem for finite, say $n$, number
of gaussian variables $(\phi_1,\phi_2,\dots ,\phi_n)\equiv\vec{\phi}^T$.
If $\vec{\phi}$ is distributed according to \footnote{Where evident 
the vector label $\;\;\vec{ }\;\;\;$
will be omitted.}
\begin{equation}
P(\phi) \sim \exp{\left( -{1\over 2} \phi^T G \phi\right) },
\end{equation}
then moments of $\phi$'s are given by the sum over all contractions
\begin{eqnarray}
 < \phi_{i_1} \phi_{i_2} \dots \phi_{i_k} > =
(\overbrace{\phi_{i_1}\phi_{i_2}}\overbrace{\phi_{i_3}\phi_{i_4}}\dots)+\\
\nonumber
(\overbrace{\phi_{i_1}\overbrace{\phi_{i_2}\phi_{i_3}}\phi_{i_4}}\dots)
+\dots\;\; , \label{wick}
\end{eqnarray}
where a single contractions reads
\begin{equation}
\overbrace{\phi_i\phi_k}=\left(G^{-1}\right)_{ik},
\end{equation}
for a non-singular matrix $G$. 
  
To match the combinatorics of contractions with that of simple permutations
required in Eq.(\ref{wei}) we introduce $2n$ gaussian variables\newline
$(\phi_1,\dots ,\phi_n),(\psi_1,\dots ,\psi_n)
=(\vec{\phi}^T,\vec{\psi}^T)\equiv\vec{\Psi}^T$ distributed 
accordingly to 
\begin{equation}
P(\Psi) \sim \exp{\left(-{1\over 2} \Psi^T M \Psi \right) },
\end{equation}
with the $2n\times 2n$ matrix
\begin{equation}
M=\left( \begin{array}{cc} 0 & A^{-1} \\
                      A^{-1} & 0 
              \end{array}  \right) ,\;\;\;
\end{equation}
and $A$ as given by Eq.(\ref{cor}).
Since, by construction, $\phi$ and $\psi$ variables are not coupled
within themselves, all non-zero contractions occur between $\psi$'s
and $\phi$'s only. Hence
it follows from the Wick theorem (\ref{wick}) that
\begin{equation}
 < \prod_i^n (\phi_i \psi_i) > = W_n(p_1,\dots ,p_n).
\end{equation} 
since $M^{-1}=\left(\begin{array}{cc} 0      & A \\
                                      A      & 0 
               \end{array} \right) $ .
This is the integral representation sought of in the Introduction.
However the choice of variables, although simple conceptually, is not
very practical, since the form $\Psi^T M \Psi$ is not positive definite,
and consequently the paths of $\prod_i^n d\psi_i d\phi_i $ integrations 
should lie in the
complex plane of the corresponding variables. 
This deficiency is removed and the integrand further simplified
by the following series of transformations of variables $(j,k=1,\dots n)$
\begin{eqnarray}
\phi_j={1\over\sqrt{2}}(u_j+i v_j), & u_j=O_{jk}\zeta_k,  
     & \zeta_j=\sqrt{\lambda_j} x_j, \\
\psi_j={1\over\sqrt{2}}(u_j-i v_j), & v_j=O_{jk}\eta_k, 
     & \eta_j=\sqrt{\lambda_j} y_j,
\end{eqnarray}
where the orthogonal matrix $O$ transforms $A$ into the diagonal 
form $O^T A O=Diag[\lambda_1,\dots,\lambda_n]$. One obtains
\begin{equation}
W_n=\int_{-\infty}^{\infty} \exp{\left(-{1\over 2}\sum_j^n(x_i^2+y_j^2)\right)}
  \prod_{\mu}^n {1\over 2}((x\cdot e^{(\mu)})^2+(y\cdot e^{(\mu)})^2) 
\prod_i^n {dx_i dy_i\over 2\pi}.
\label{xy}
\end{equation}
Vectors $\vec{e}^{(\mu)}$ are given by 
\begin{equation}
e^{(\mu)}_i=O_{\mu i}\sqrt{\lambda_i}, \label{emu}
\end{equation} 
and all integrals are along the real axis.
It turns out that all eigenvalues of the correlation matrix, Eq.(\ref{cor}),
are nonnegative. The more general case of arbitrary eigenvalues can be readily
dealt with by the appropriate Wick rotations which will be different 
for different coordinates.

   Equation (\ref{xy}) represents the sum of $n!$ terms in (\ref{wei})
by the $2n$-fold integral of the gaussian type. The whole dependence 
on the original momenta $(p_1,\dots , p_n)$ is contained in vectors 
$\vec{e}^{(\mu)}$
which, in view of Eq.(\ref{emu}), are closely related to the eigenvectors 
of the correlation matrix A. 

   The integral (\ref{xy}) can be calculated by the Monte Carlo
technique. We have found, however, that the importance sampling according, 
to the normal distribution, is not sufficient to produce a decent estimate.
The product of $n$ factors dramatically modifies the relatively mild gaussian 
dependence and should be included in the weight. We have therefore solved
the integral analytically in the saddle point approximation. This gives
rather satisfactory results as discussed in the next Section. 
\section{Saddle point approximation}
\subsection{The action and the saddle point equations} 
The interplay between the gaussial fall off and the polynomial rise
of the two factors in the integrand of Eq.(\ref{xy}), produces
a maximum far from the origin. This maximum becomes more
and more narrow with increasing $n$. Thus, we expect that the saddle point
approximation will be better for higher $n$, just where the exact
summation (\ref{wei}) fails in practical terms. 

If one  writes the integrand in Eq.(\ref{xy}) as
\begin{equation}
I(x,y)={1\over (4\pi)^n} \exp{(-S(x,y))},
\end{equation}
then the "action" $S(x,y)$ reads
\begin{equation}
S(x,y)={1\over 2} (\vec{x}^2+\vec{y}^2)-\sum_{\mu}^n\ln{(x_{\mu}^2+y_{\mu}^2)},
\label{act}
\end{equation}
where the greek subscript denotes the scalar product 
$z_{\mu}\equiv(\vec{z}\cdot \vec{e}^{(\mu)}),\;\;z=x,y$. 
Then the location of the saddle point (or points) is determined by
the following system of nonlinear equations
\begin{eqnarray}
\vec{x} & = & 2\sum_{\mu}^n \vec{e}^{(\mu)} 
{ x_{\mu}\over (x_{\mu}^2+y_{\mu}^2)  } ,  \label{sad1} \\ 
\vec{y} & = & 2\sum_{\mu}^n \vec{e}^{(\mu)} 
{ y_{\mu}\over (x_{\mu}^2+y_{\mu}^2) }.
\label{sad2}
\end{eqnarray}
In general many  solutions may exist and there is no
simple method to find all of them. The problem is partially alleviated
by identifying the continuous symmetries of the integrand. This
is also necessary for the proper application of the saddle point
technique. Therefore we will first discuss the symmetries of the problem
and identify corresponding zero mode. 
\subsection{Global U(1) symmetry and gauge fixing}
The integrand in Eq.(\ref{xy}) is invariant under the "horizontal"
U(1) rotation
\begin{eqnarray}
\vec{x}(\alpha)&=&\phantom{-}\vec{x} \cos{(\alpha)} + \vec{y} \sin{(\alpha)}, 
\\ 
\vec{y}(\alpha)&=&-\vec{x} \sin{(\alpha)} + \vec{y} \cos{(\alpha)},
\label{sym}
\end{eqnarray}
with the same $\alpha$ for every coordinate $x_i,y_i$.
Consequently there is no unique solution to Eqs.(\ref{sad1}-\ref{sad2}) and 
one has to fix this freedom before the saddle point technique
can be used. We follow the standard procedure known from the 
 field theory. In this language the symmetry (\ref{sym})
is the global one, and the single "gauge fixing" condition is required.
We choose the symmetric gauge  
\begin{equation}
\vec{x}^2=\vec{y}^2, \label{gfix}
\end{equation}
and insert the following representation of unity under the integral 
(\ref{xy})
\begin{equation}
\Delta(x,y)\int_0^{2\pi} \delta(x(\alpha)^2-y(\alpha)^2) d\alpha=1. \label{fix}
\end{equation}
The gauge invariant function $\Delta(x,y)$ follows from Eq.(\ref{fix}) 
\begin{equation}
 \Delta(x,y)={1\over2}(\vec{x}^2+\vec{y}^2). \label{del}
\end{equation}
\subsection{The formula}
With the aid of Eqs.(\ref{fix}) and (\ref{del}) the volume of the gauge orbit
$(2\pi)$ can be readily factored out and the zero mode corresponding to the 
symmetry (\ref{sym}) becomes fixed. 
Further calculations are straightforward and the final result reads
\begin{equation}
W_n=2^{1-n}|s_0|\sqrt{\pi\over{{\prod^{'}}_i^{2n} \Lambda_i}} 
\exp{(-S(s_0,s_0))},
\label{fin}
\end{equation}
where the product is taken over all, but one (equal to zero), 
eigenvalues of the 2n
 dimensional covariance matrix. 
\begin{equation}
S^{(2)}_{jk}={ \partial^2 S(x,y)\over\partial z_j\partial z_k }|_{(s_0,s_0)}, 
\end{equation}
evaluated at the saddle point $(\vec{s}_0,\vec{s}_0)$. 
For completness we quote the explicit form of $S^{(2)}$ for arbitrary 
$x$ and $y$.
\begin{equation}
S^{(2)}(x,y)=\left( \begin{array}{cc}
                        S^{(xx)} & S^{(xy)}  \\
                        S^{(yx)} & S^{(yy)} 
                   \end{array}  \right),
\end{equation}
\begin{eqnarray}
S^{(xx)}_{ik}&=&\delta_{ik}-2\sum_{\mu}^n e^{(\mu)}_i e^{(\mu)}_k
{y_{\mu}^2-x_{\mu}^2\over(x_{\mu}^2+y_{\mu}^2)^2}, \\
S^{(xy)}_{ik}=S^{(yx)}_{ik}&=&4\sum_{\mu}^n e^{(\mu)}_i e^{(\mu)}_k
{x_{\mu} y_{\mu}\over(x_{\mu}^2+y_{\mu}^2)^2}, \\
S^{(yy)}_{ik}&=&\delta_{ik}-2\sum_{\mu}^n e^{(\mu)}_i e^{(\mu)}_k
{x_{\mu}^2-y_{\mu}^2\over(x_{\mu}^2+y_{\mu}^2)^2}. 
\end{eqnarray}
Condition (\ref{gfix}) does not entirely fix the freedom (\ref{sym}).
 For each symmetric solution, $(\vec{s},\vec{s})$ say,
there exist also three rotated solutions $(-\vec{s},\vec{s})$
$(\vec{s},-\vec{s})$ and $(-\vec{s},-\vec{s})$ which belong to the
orbit (\ref{sym}) and satisfy (\ref{gfix}). Of course all four (equal) 
contributions are included in the final result. Chosing different
gauge may lead to the different discrete degeneracies, but 
the final result is not changed. More important, we have found that,
except for the above degeneracy, there is no {\em other, essentially
different,} solution of the saddle point equations. This seems to be a
general property of our $S(x,y$) which was checked for a number of
choices of correlated momenta $(p_1,\dots ,p_n)$. 

Finally, a few comments about the numerical solution of the system
(\ref{sad1}-\ref{sad2}). Since our experiments indicate that after a 
complete gauge fixing
there is only a single maximum, any numerical routine should perform satisfactorily.
We have used a simple iterations of Eqs.(\ref{sad1}-\ref{sad2}) which quickly
converged to the cycle between two parallel vectors. Then, a repeated 
mixture
of the arithmetic average between the two followed by one iteration of the map
(\ref{sad1}-\ref{sad2}) converged rapidly to the unique solution. Condition
(\ref{gfix}) was assured by the choice of the symmetric starting
point.   

\section{Results and further applications}

Table I contains comparison between the saddle point formula, 
Eq.(\ref{fin}), and the exact one (\ref{wei}). As expected, the accuracy 
of the saddle point calculation increases with increasing $n$ reaching
the level of 2\% at $n=8-9$. Consequently, both methods applied complementarily
would give a satisfactory approach for all $n$.

Results presented in Table I were obtained for the one dimensional
momenta generated randomly from the interval $0< p_k < 2$ and with
the width in Eq.(\ref{cor}) $\sigma=1/\sqrt{2}$. This corrresponds
to very strongly correlated particles. In real applications,
few particles are clustered within a range of typically measured
$\sigma\sim 0.1 GeV$. This situation corresponds to the block diagonal
(after suitable permutation) form of matrix $A$ - each block 
describing one cluster of identical bosons. It can be readily seen
that, for exactly block diagonal (reducible) A, the integral
(\ref{xy}) factorizes
\begin{equation}
W_n=W_{n_1} W_{n_2} \dots W_{n_{Ncluster}}.
\end{equation}
This factorization follows even more clearly from the definition
(\ref{wei}). On the other hand the saddle point formula applied
directly to the whole matrix $A$ would fail since it does not
accomodate multiple zero modes (one for each cluster) which
occur in the reducible case. The remedy of this difficulty is 
evident. One should first perform reduction of a given event
into irreducible clusters and then, depending on the size of the cluster,
apply either exact or saddle point formula to each block separately.
The reduction procedure will contain an essential parameter, say $\epsilon$,
which can be defined as a tolerance beyond which bosons would be considered
 uncorrelated. This tolerance cannot be too small since then a
quasi block diagonal matrix would be qualified as a single irreducible
cluster. Therefore the value of epsilon is determined by the performance
of the saddle point formula for weakly broken additional U(1) symmetries.
Our numerical experiments indicate that the present "one loop" formula
works for $\epsilon > 0.05 - 0.1$.

    Further improvement in the accuracy of the saddle point result
can be easily achieved by calculating higher corrections. Including
for example the two additional terms of the Taylor expansion around 
the saddle
should not increase much the computational effort but would reduce 
the errors considerably. More important, this should allow to decrease
 the tolerance parameter $\epsilon$ discussed above. 

   Another path for an improvement is the Monte Carlo calculation of the
corrections to the saddle point formula. As mentioned Section 2, 
MC calculation of the average (\ref{xy}) was not satisfactory, because
the samples were drawn from the wrong distribution (centered gaussians).
However one can use the saddle point approach to locate the maximum
of the integrand and then use the saddle point approximation of the integrand
as the proper distribution of generated samples. In that case the observable
(the ratio between exact and saddle point integrands) will be a slowly 
varying function
of $\vec{x}$ and $\vec{y}$ and statistical estimates will be reliable.

   Finally, there exists an intriguing possibility of defining a new family
of event generating MC programs which would have the corrections (\ref{wei})
included on the stochastic basis. That may be possible since the integrand
in our representation (\ref{xy}) is {\em positive}. Therefore it might be 
feasible to modify existing event generators by incorporating that factor.

To summarize, we have derived an exact integral representation
for the Bose-Einstein weights, which until now have been computed
by summing over $n!$ terms. Resulting $2n$-dimensional integral
was calculated by the saddle point technique, and obtained
formula approximates the exact sum to less than 2\% for more than 7
correlated bosons. Various applications and improvements have been outlined.
\vspace*{.4cm}

I would like to thank K. Fia\l kowski for the discussion.
This work is supported in part by the Polish Committee for Scientific
Research under the grant no. 2P03B19609.
 \newpage

  \begin{table}    
  \begin{center}
   \begin{tabular}{|c|c|c|c|c|c|c|} \hline\hline
{\em n} & $W_n^{saddle}$ & $t_{saddle} (sec)$ & $W_n^{exact}$ & $t_{exact} (sec)$ & error (\%) &  $a$\\ \hline\hline
  2     & 1.000  & 0.4         & 1.125  &         &  12.5 & 0.71    \\ \hline
  3 &   1.878   &  0.6      & 2.094  &          &  10.3 & 0.68   \\
    \hline
  5 &   13.144  &  1.2      & 14.196  &  0.1   &   7.4  &   0.64 \\
    \hline
  7 &   649.01  &  2.0      & 666.78  &  5.0   &   2.7  &   0.74 \\
      \hline
  8 &   8628.2  &  2.5      & 8748.2  &  44.7   &   1.3  &  0.82 \\
    \hline
  9 &  43688.   &  3.2      & 44391.  &  443.  &   1.5  &  0.79 \\ 
    \hline\hline
    &           &        &  n!     &        &        &        \\
    \hline
  10 & 5.223e04 &  3.9   & 3.63e06 &       &         &  0.65  \\
    \hline
  15 & 2.216e09 &  7.0   & 1.31e12 &       &       &  0.65 \\
  \hline
  20 & 1.034e15  &  11.4 & 2.42e18 &          &    & 0.68 \\
    \hline
  25 & 2.557e22  &  16.4 & 1.55e25 &        &      & 0.77 \\ 
    \hline
  30 & 3.118e26  &  22.5 & 2.65e32 &          &    & 0.63\\
    \hline
  35 &  4.112e32 &  30.1 & 1.03e40 &          &    & 0.61\\ 
    \hline
  40 &  3.449e39 &  36.0 & 8.16e47  &          &    & 0.62\\
   \hline\hline  
   \end{tabular}
  \end{center}
\caption{Saddle point results and the exact summation for
different number $n$ of identical, strongly correlated, bosons.
Timing (in seconds) of the Mathematica runs on a HP735 
workstation are also quoted. The last column gives the result of a 
simple fit $W_n=n!a^n$, where the parameter $a^2$ is a measure of the average
value of the matrix elements of $A$. }
   \end{table}


\begin{thebibliography}{99}
\bibitem{HBT} R. Hambury-Brown and R. Q. Twiss, Nature {\bf 178} (1956) 1046.
\bibitem{GG} G. Goldhaber, S. Goldhaber, W. Lee and A. Pais, Phys. Rev.
{\bf 120} (1960) 300.
\bibitem{PR} S. Pratt, Phys. Rev. Lett. {\bf 53} (1984) 1219.
\bibitem{HA} S. Haywood, Rutherford Lab. report, RAL 94-07 (1995).
\bibitem{BK} A. Bia\l as and A. Krzywicki, Phys. Lett. {\bf B354} (1995) 134;
 {\em see also}\newline  A. Bia\l as, Jagellonian Univ. preprint, 
 TPJU-20/96, to
appear in the Proceedings from the 7th International Workshop on Multiparticle
Production, Nijmegen, June 30 - July 6, 1996.   
\bibitem{SJ} L. L\"{o}nnblad and T. Sj\"{o}strand, Phys. Lett. {\bf B351}
  (1995) 293.
\bibitem{FW} K. Fia\l kowski and R. Wit, Zeit. Phys.{\bf C}, {\em in print}.
\bibitem{ZJ} S. Jadach and K. Zalewski, private communication.
\end{thebibliography}
\end{document}